\documentclass{article}
\pdfoutput=1

\usepackage{arxiv}

\usepackage[utf8]{inputenc}
\usepackage[T1]{fontenc}
\usepackage{hyperref}
\usepackage{url}
\usepackage{booktabs}
\usepackage{amsfonts}
\usepackage{nicefrac}
\usepackage{microtype}
\usepackage{graphicx}
\usepackage{doi}
\usepackage[margin=0.5cm]{subcaption}
\usepackage{float}

\RequirePackage[labelfont={bf},%
                labelsep=period,%
                justification=raggedright]{caption}

\title{An Orchestration Platform that Puts Radiologists in the Driver's Seat of AI Innovation: A Methodological Approach}

\author{ \href{https://orcid.org/0000-0002-2251-6740}{\includegraphics[scale=0.06]{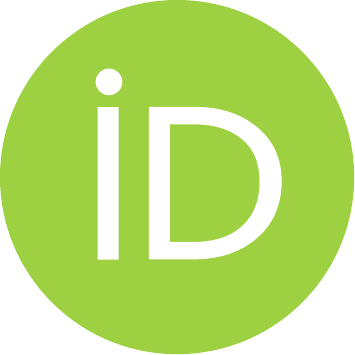}\hspace{1mm}Raphael Y.~Cohen}\\
	Division of Emergency Radiology\\
	Department of Radiology\\
	Brigham and Women's Hospital\\
	Boston, MA 02115 \\
	\texttt{rcohen15@bwh.harvard.edu} \\
	\And
	\href{https://orcid.org/0000-0003-1275-681X}{\includegraphics[scale=0.06]{orcid.pdf}\hspace{1mm}Aaron D.~Sodickson} \\
	Division of Emergency Radiology\\
	Department of Radiology\\
	Brigham and Women's Hospital\\
	Harvard Medical School\\
	Boston, MA 02115 \\
	\texttt{asodickson@bwh.harvard.edu} \\
}

\renewcommand{\shorttitle}{An Orchestration Platform that Puts Radiologists in the Driver’s Seat of AI Innovation}

\hypersetup{
pdftitle={An Orchestration Platform that Puts Radiologists in the Driver's Seat of AI Innovation: A Methodological Approach},
pdfsubject={eess.IV, cs.AI, cs.DC},
pdfauthor={Raphael Y.~Cohen, Aaron D.~Sodickson},
pdfkeywords={Medical Imaging, Artificial Intelligence, Distributed Systems},
}

\begin{document}
\maketitle

\begin{abstract}
Current AI-driven research in radiology requires resources and expertise that are often inaccessible to small and resource-limited labs. The clinicians who are able to participate in AI research are frequently well-funded, well-staffed, and either have significant experience with AI and computing, or have access to colleagues or facilities that do. Current imaging data is clinician-oriented and is not easily amenable to machine learning initiatives, resulting in inefficient, time consuming, and costly efforts that rely upon a crew of data engineers and machine learning scientists, and all too often preclude radiologists from driving AI research and innovation. We present the system and methodology we have developed to address infrastructure and platform needs, while reducing the staffing and resource barriers to entry. We emphasize a data-first and modular approach that streamlines the AI development and deployment process while providing efficient and familiar interfaces for radiologists, such that they can be the drivers of new AI innovations.
\end{abstract}


\flushbottom

\thispagestyle{empty}

\section{Background}
When our small Emergency Radiology lab sought to engage in AI research, we found that we lacked needed resources, and pre-existing AI research systems did not translate to our workflow or adapt to our needs. Without a system to manage the many facets of setting up and performing AI research, significant manual efforts and a constellation of incongruent tools are needed. A wide range of effort-intensive operations combined to make AI research infeasible for us: Data curation, annotation, machine learning model development, management of people and resources, security, auditing, and multi-system interoperability are far too large of a simultaneous undertaking for a resource-limited lab to manage. The costs of a large staff and requisite resources to perform all of these activities were prohibitively high.

In order to perform rapid research, development, and deployment of AI models with minimal staff and low-cost resources, we needed a system that could orchestrate all of these necessary tasks, without the omissions, gaps, and incongruities between tools that so often require many resources and manual intervention. We set out to design an integrated platform that could facilitate the plurality of our research initiatives. Our goal was to restore radiologists as the drivers of innovation in imaging-focused AI. Our design philosophy was that tasks that could be automated, such as handling, translating, and curating high-quality data, should be handled by computers rather than armies of annotators, data scientists, and engineers.

The hurdles to successful facilitation of imaging machine learning have been well documented \cite{PreparingMedicalImaging}. Non-standardized and non-normalized data present difficulties for data ingestion and data annotation. The DICOM standard was instrumental in allowing the creation of PACS and the digitization of imaging records \cite{DICOM}. The subsequent development of DICOM-SR to incorporate structured reports further expanded the scope of the DICOM format \cite{DICOMSR}. However, at present, different scanners and vendors often implement DICOM differently, and hospitals may use different approaches to enter data into similar metadata fields. With many optional tags that vary from vendor to vendor, and with varied practices in recording data between hospitals, imaging data in the DICOM format is often still varied and highly inconsistent. 

Annotation is the most time consuming portion of the development process, and is an expensive task for radiologists to perform. If data is in a format that prevents successful ingestion, data engineers may be required to intervene to attempt to salvage radiologists' annotations, or annotations may need to be discarded. If data ingestion produces corrupted or incorrect data and these data are included in model training, they can threaten the integrity of the model results.

Many groups have developed systems that target components of the AI development pipeline to reduce the workload needed to facilitate these research efforts. These include the development of faster annotation tools for radiologists  \cite{DicomAnnotator, ePad}, workflows that automate data organization for data scientists \cite{DVC}, and approaches that can speed up model exploration \cite{AutoML}. Many groups have leveraged advances in cloud-computing and have drawn from innovations in the development of PACS as inspiration for building platforms that automate portions of the machine learning research workflow \cite{Prevedello, VNA_AI}. In recent years, the adoption of active machine learning strategies have offered reductions in both annotation time and model training efforts\cite{Prevedello, RILContour, DeepInfer}. In these approaches, an initial annotated dataset is used to train preliminary models, which are then iteratively updated and used to propose annotations on remaining images so that radiologists need not start from scratch, and may begin instead by refining the model's proposed annotations. Systems like RIL-Contour from the Mayo Clinic have noted a nearly 80\% decrease in annotation time with the use of these strategies.

The driving force behind the development of our platform and quest for automation was the core problem for our small lab: People are the most expensive resource in the AI development process, especially radiologists. Our requirement was to build a platform that minimized those costs by shifting as much work as possible to be handled by computer systems. In order to accomplish this, data had to be translated to be amenable for machine analysis as early as possible in the workflow, and all subsequently built modules had to be highly optimized for their use.

\begin{figure}[ht]
\centering
\includegraphics[width=0.8\linewidth]{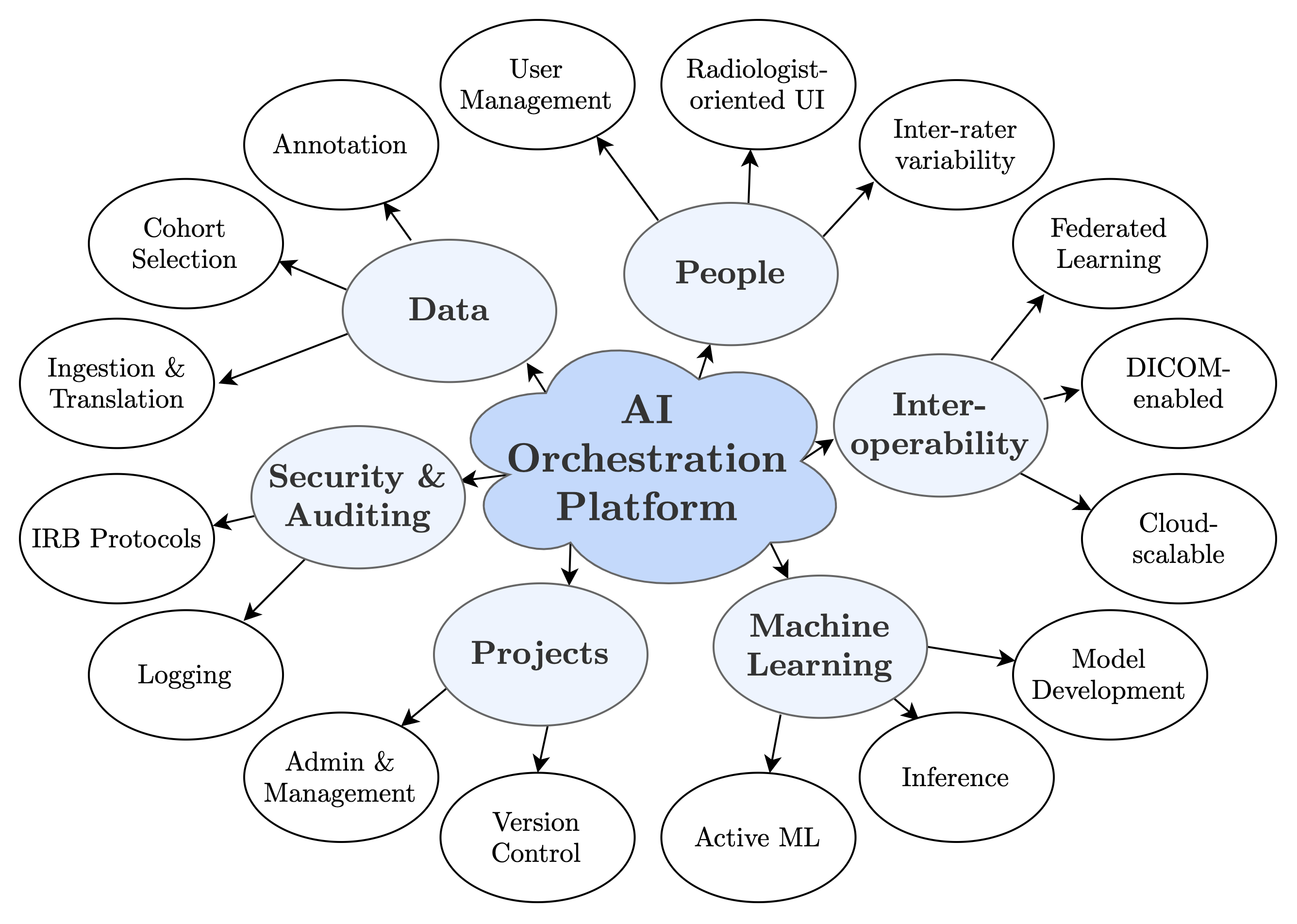}
\caption{Functionality Schematic of Our AI Orchestration Platform}
\label{fig:tomosuite_functions}
\end{figure}

The system we built includes the functionalities shown in Figure \ref{fig:tomosuite_functions}. Numerous obstacles needed attention and resolution to achieve the necessary high degree of automation, most notably, issues with varied, messy, and inconsistent data. Our platform is designed as the composition of modular components. It is able to scale and grow with resource needs; it is capable of handling projects pursued by small labs or larger institutions, and can even facilitate federated learning approaches.

We will first describe key decisions that were critical in the development of our platform. We will then present the platform architecture, highlighting important components and workflow design. Finally, we will discuss key benefits of our implementation in practical AI research.

\clearpage

\section{Methods}
The design of our system eliminated as much manual intervention as possible to reduce the human effort needed to complete projects. Curating high-quality and machine-learning ready data from the outset were key design principles to accomplish this goal. Translating and transforming data to be continuously handled by computer systems under the hood, while maintaining familiar workflows for radiologists, provided a backbone that facilitates AI research efforts with minimal barriers to completion. The system architecture was designed to run both on low-cost hardware and on large infrastructure like Amazon Web Services (AWS) or Microsoft Azure, in a manner that is proportional to resource needs, available funding, and research staff.

At the outset, two general areas of focus emerged while analyzing the cost of different types of failures and bottlenecks throughout the development lifecycle: Constructing a pipeline framework that can be fully automated, and creating components, such as an annotation platform, that are highly efficient in their function to minimize time spent on their usage. Addressing these two areas was critical in guiding the creation of all parts of the overall system.

\subsection{Key Design Principles for a Highly Automated Pipeline}
\subsubsection*{Workflow Organization}
\begin{figure}[H]

\begin{subfigure}[t]{\linewidth}
\centering
\includegraphics[width=0.9\linewidth]{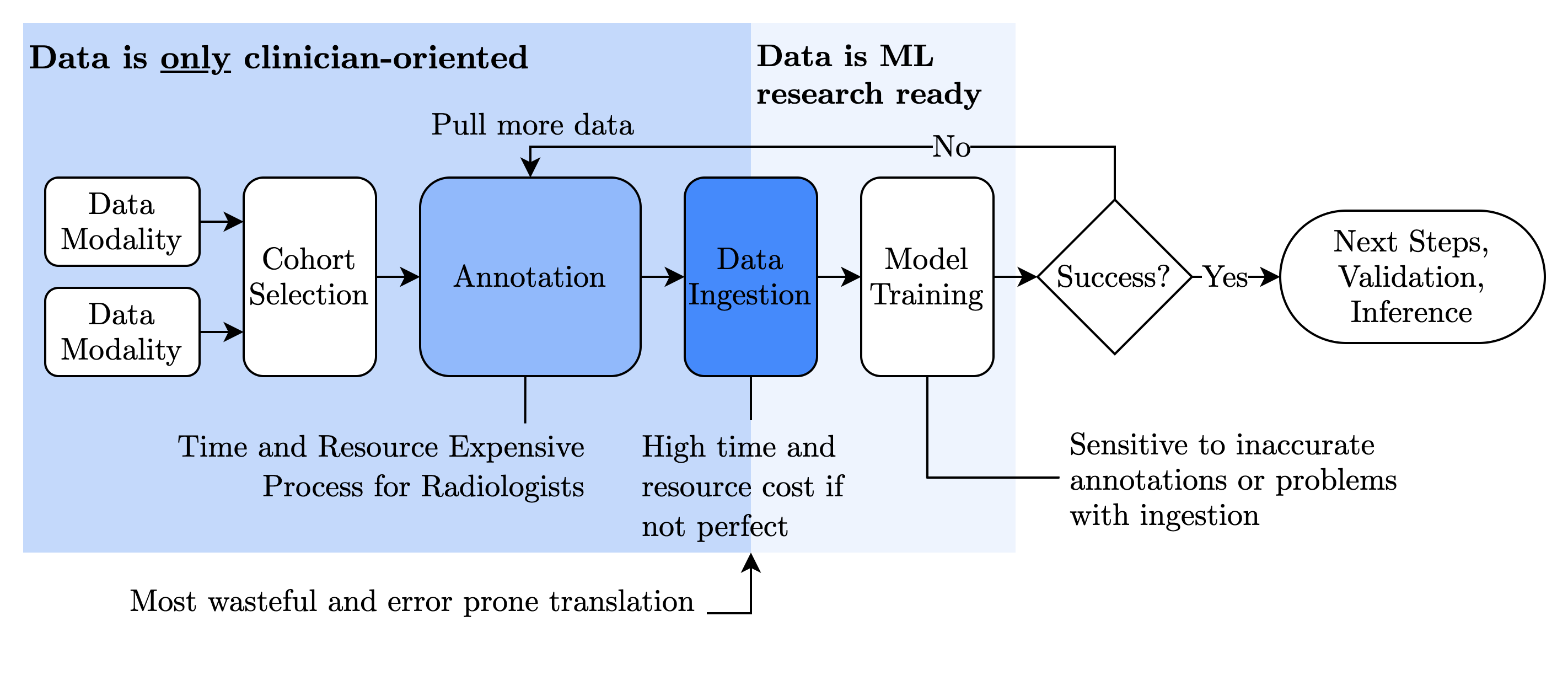}
\subcaption{Traditional ML Workflow}
\label{fig:ai_workflow_old}
\end{subfigure}
\begin{subfigure}[t]{\linewidth}
\centering
\includegraphics[width=0.9\linewidth]{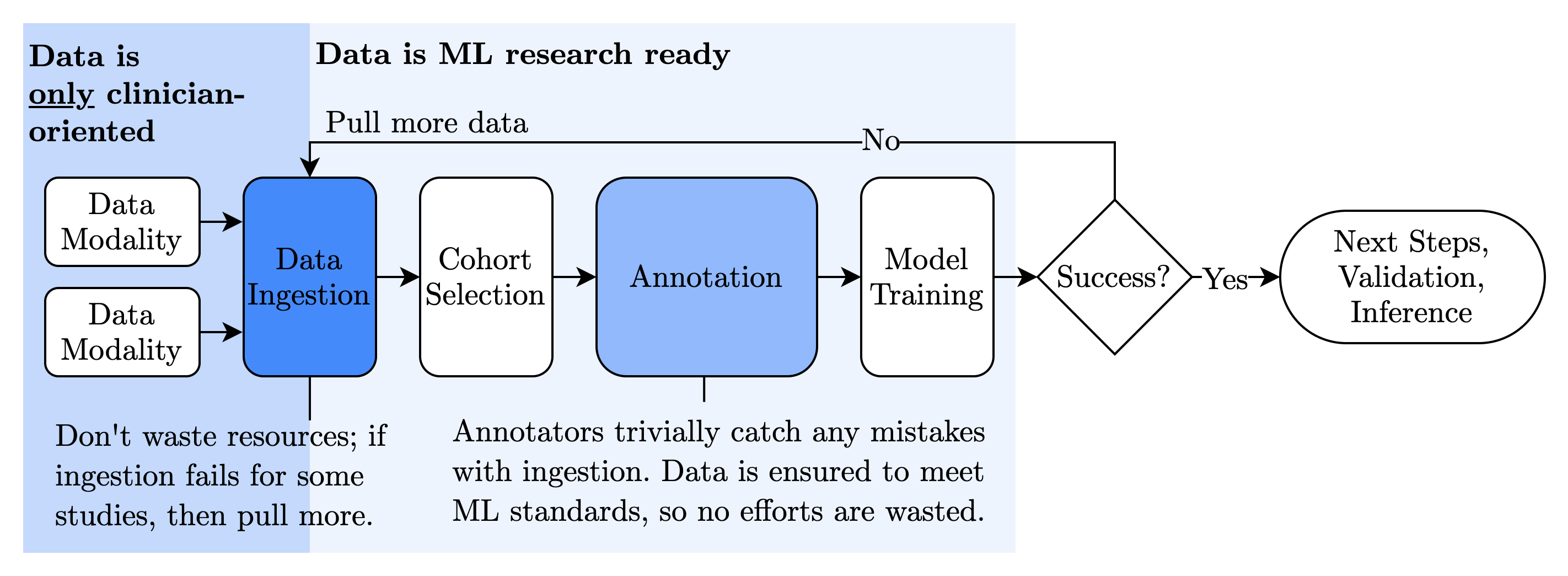}
\subcaption{Waste-Minimizing Workflow}
\label{fig:ai_workflow_new}
\end{subfigure}

\caption{\textbf{(a)} Traditional machine learning development workflow. \textbf{(b)} Workflow designed to minimize resource waste by translating to ML-ready data early. Annotation data must be saved in a manner that ensures clean and standardized metrics are met. Data cleaning tasks are thereby minimized as data ingestion no longer has to contend with salvaging annotations on malformed images. Other challenges with data that previously may have gone unnoticed can be recognized early and upfront, before they threaten model training.}
\label{fig:ai_workflow}
\end{figure}

The traditional order of ML Workflow operations is shown in Figure \ref{fig:ai_workflow_old}, whereas our modified sequence is shown in Figure \ref{fig:ai_workflow_new}. The modified workflow is designed to conserve all annotation efforts by placing data annotation \textit{after} data ingestion, in order to position the most expensive part of the pipeline after the most error-prone part. As data is ingested, image volumes are converted into matrix object representations. Associated metadata fields are placed in a database, where common metadata are resolved to uniform representations and structures. Data is then maintained in a machine-amenable format throughout the entire pipeline, eliminating the need for data engineers at any stage. This reorganization provides the additional benefit that annotators can trivially spot ingestion errors in imaging data that might otherwise have progressed unnoticed to model training.

\raggedbottom

\subsubsection*{Structure Data to be Anonymized with Ease}
In addition to the matrix representation of image data and the standardization of metadata tags in storage, the platform separates metadata storage between Protected Health Information (PHI)-containing data and non-PHI data. All components are built to run without PHI data, and the system requires that specific access be granted for users to receive non-anonymized data. The platform uses an inclusionary list of metadata fields verified to be free of any PHI, rather than an exclusionary list, to ensure that PHI data cannot be unintentionally sent to users who are not designated to receive it. The only potential risk of PHI inclusion is thus burned-in data on the images themselves, which would require special handling.

\subsubsection*{Keep Development Simple to Reduce Cross-Dependencies and Bottlenecks}
\begin{figure}[ht]
\centering
\includegraphics[width=\linewidth]{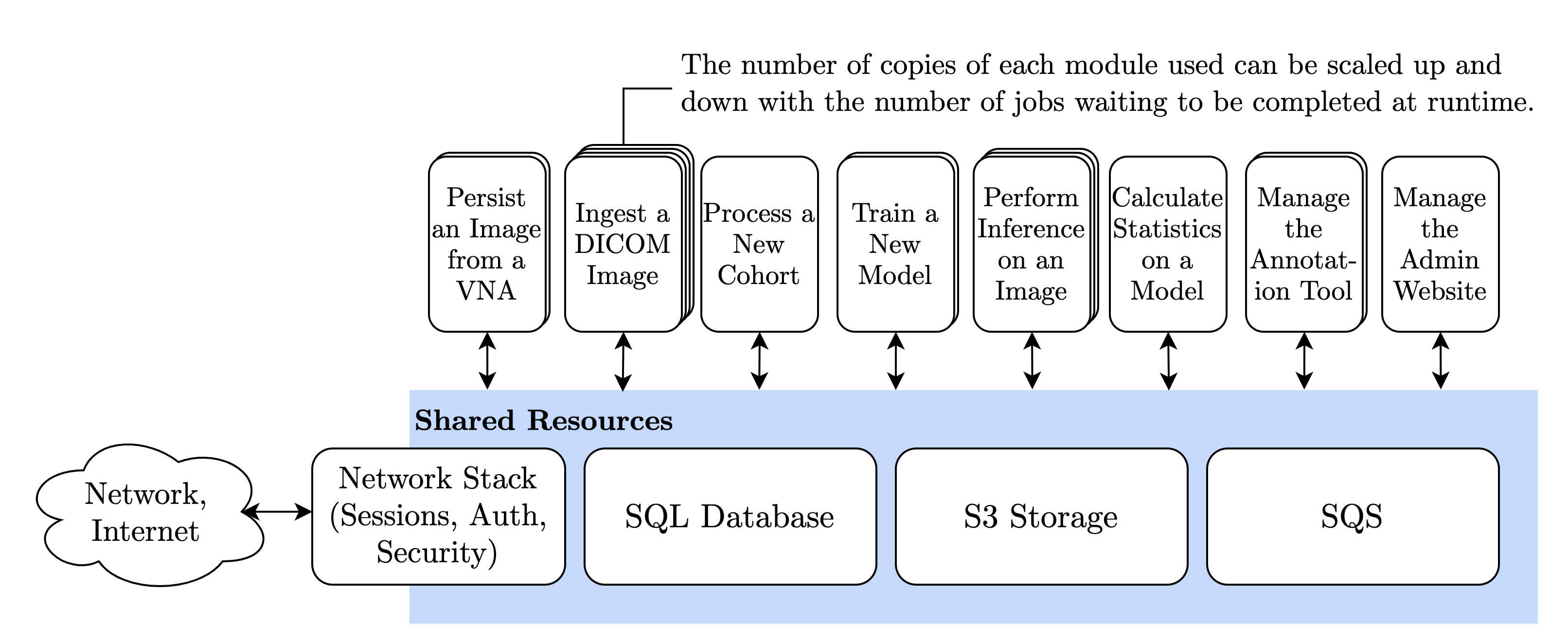}
\caption{Multiple modules working in concert without ever directly communicating with each other. Jobs are posted to SQS queues, and modules read from them. Modules are containerized with Docker and managed by Kubernetes, so that the platform can scale automatically to directly respond to the needs of users. Modules shown are examples, and not exhaustive of practice.}
\label{fig:modular_workflow}
\end{figure}

The platform architecture is based on simple modules that perform individual tasks and do not have cross-module dependencies that can create bottlenecks in development and execution. This methodology is intrinsically optimized for cloud-computing resources and multi-system distributed workloads. As shown in Figure \ref{fig:modular_workflow}, all modules perform one and only one task: Module algorithms are developed to strictly convert predefined, structured input into predefined, structured output, independent of all other modules in the pipeline. Importantly, no module is allowed to directly communicate with any other module in the system, other than to three storage resources, or the networking stack when needed.

\clearpage

The platform utilizes three storage resources:
\begin{itemize}
    \item PostgreSQL Database: Storage of relational data, metadata, or other structured data relevant to images, cohorts, or projects.
    \item Simple Storage Service (S3): A file-like storage solution that is standard to cloud computing technologies like AWS. The platform uses this for any large storage object, such as image-matrix objects, annotation matrices or vectors, or ML models.
    \item Simple Queue Service (SQS): Another cloud computing technology common to vendors like AWS. This is most akin to a messaging board, where modules look for jobs they can perform, and post their results once complete.
\end{itemize}

This rigorous modularity and design framework takes full advantage of advances in cloud computing and virtualization technologies. The many modules of the platform server are all containerized with Docker. The number of copies of each container is managed by Kubernetes and can be dynamically scaled up and down to meet the workflow needs in real-time. The networking stack is similarly containerized and individually isolated by functionality, such as managing connections, authentication, and security, to allow for easy management, upgrading, and even replacement as needed. Maintaining one path in and out for data entering or leaving the system enables easy management of access and security.

\subsubsection*{Parallelize Everything}
\begin{figure}[ht]
\centering
\includegraphics[width=0.6\linewidth]{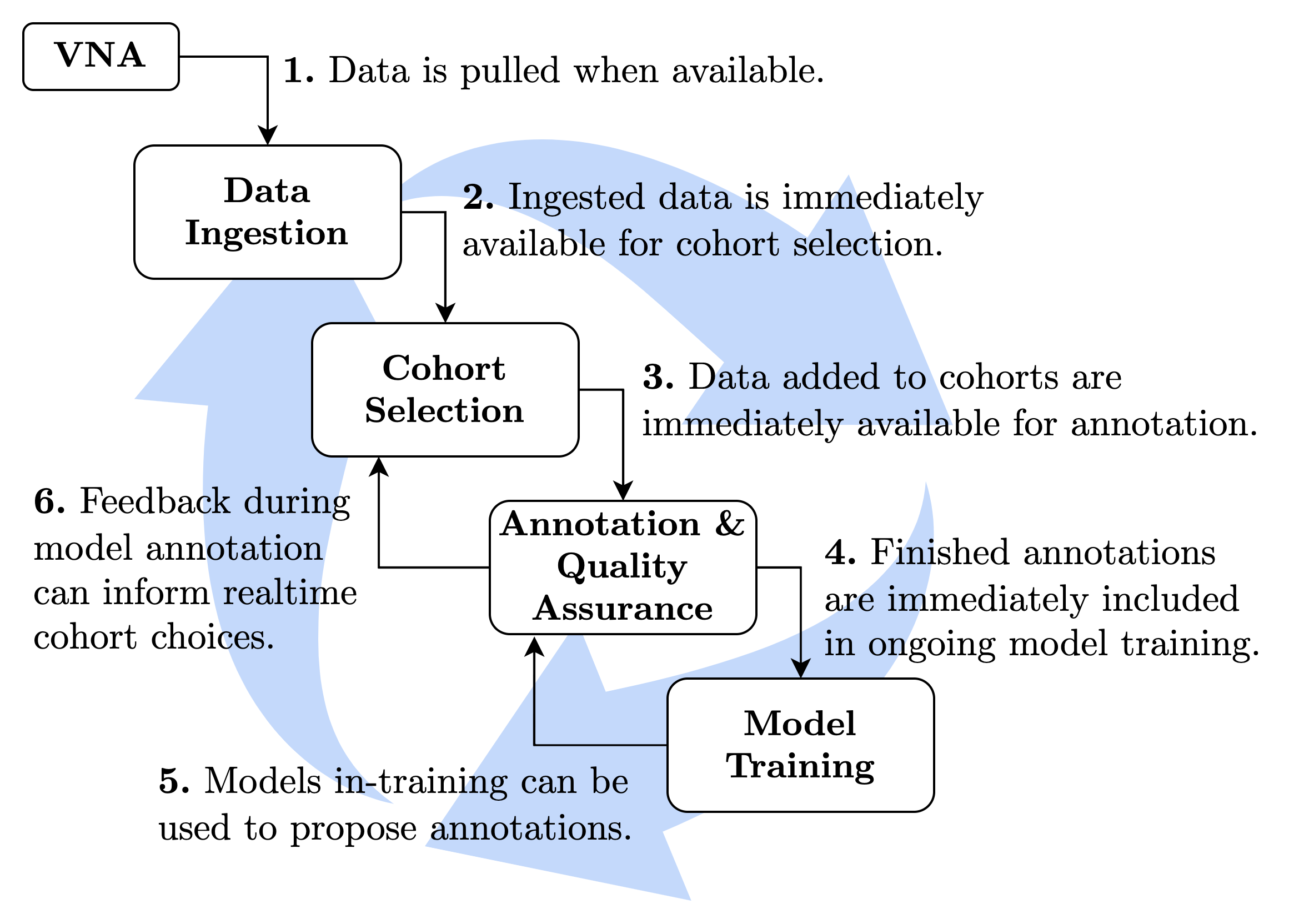}
\caption{Parallelized Workflow Utilizing an Active Machine Learning Strategy}
\label{fig:parallel_workflow}
\end{figure}

The modular structure enables the parallelization of many facets of the machine learning workflow, further enabling faster result turnaround, as shown in Figure \ref{fig:parallel_workflow}. Data Ingestion, Cohort Selection, Annotation, Quality Assurance, Model Training, and subsequent evaluation and response need not be serial tasks completed in full before moving on to subsequent steps. Instead, the pipeline is responsive to data as it becomes available. As data is ingested, it is immediately available for cohort selection. As data enters into a cohort, radiologists are immediately able to begin annotation. As annotations are completed, they are immediately available for model training. This allows all stages of the process to proceed in parallel, eliminating common workflow bottlenecks in the AI development pipeline.

\subsection{Creating Highly Efficient Tools in the AI Development Pipeline}

The framework above enables a pipeline that can be automated from end to end, but the individual tools and components with which radiologists and data scientists interact must be optimized to facilitate their efficient use. These applications and workflows are best categorized into optimized annotation software, people and project management software, model development and evaluation tools, and solutions to access data and models, or integrate with other systems.

\subsubsection*{Designing an Efficient Annotation Tool for Radiologists that Maintains Structured, Machine Learning-Ready Data}
The annotation solution should accomplish three aims: to feel natural for experienced radiologists to use; to accelerate the creation and curation of high-quality annotated data; and to be harmonious with the ML-oriented ecosystem that facilitates the data that machine learning scientists need.

To achieve this, the annotation software uses the ML-ready data on the backend and faithfully reproduces the user experience of a PACS. Data and annotations are stored and handled as matrices and vectors, even behind the scenes in the annotation software itself. This means that no free-text annotations are permitted. To create the user-friendly interface on-the-fly, these data are combined with project-level configuration files that instruct the program how to translate vector data into human-readable values and familiar controls. 

Radiologists were integral to the design of the software functionality as well as the features that allow them to accomplish their annotation tasks as efficiently as possible. A key component of this workflow optimization is to minimize clicks and other interactions. Annotation categories include study, series, and slice-level labels, bounding boxes, and voxel masks. All region of interest segmentations may also have associated labels. Tools to enable efficient segmentation include voxel range masking, inclusion and exclusion masks, multi-slice painting, and morphological transformations for voxel masks (erosion, dilation, opening, closing).

The system facilitates multi-person annotation efforts, and enables quality assurance and review. Integrated tools can compare annotations for inter-rater variability, evaluate annotator performance, provide messaging and notes between annotators, and display differences between less-experienced and seasoned annotators as a teaching tool.

To adhere to the strict requirements of an ML-ready pipeline, the annotation platform prevents users from signing off on incomplete work, and enforces that all data is machine learning-ready at all stages of development. Project leads create standardized annotation templates for each project and the system enforces that these templates are completed for each annotation. Cohort management tools allow radiologists to see an overview and status of studies included in the project, with the ability to add other relevant cases to the annotation pool with ease. Data and actions are logged, version-controlled, and tracked by the system.

\subsubsection*{Creating a Portal to Manage Data, People, Projects, Models, and Inference}
The platform includes management tools for all the resources needed in AI development. Website portals include a management portal and an administrative portal. The management portal can present statistics and overview information; manage IRB Protocols; create, manage, and track users; acquire and version-control data; develop and track project progress; evaluate and deploy machine learning models; and generate usage, logging, and auditing reports of all components within the platform ecosystem. In managing data security, the management portal includes the controls necessary to associate users, data, projects, and models with IRB protocols as required by the backend in order to strictly limit access to those with appropriate credentials.

\subsubsection*{Integrating Model Training to Run Natively}
A built-in model zoo of available networks allows users to choose models with which to begin training. Architecture and model parameters can be selected by a data scientist or AI-oriented user at the onset of model training. Data scientists can upload their own model code in adherence with a predefined API. The system automatically begins training as soon as data is available.

\subsubsection*{Leveraging Active Machine Learning}
The platform integrates an active machine learning strategy, saving copies of well-performing models as they train, and using them to propose annotations. In the parallelized workflow, the active machine learning model continues to train on new annotations, which are automatically added to the training cohort as soon as they are completed. The platform evaluates the active ML models according to metrics that can be pre-selected or defined by radiologists or data scientists, with thresholds for when they consider the active ML model 'mature enough' to use for annotation proposal. When an annotator requests an annotation proposal, the highest-fidelity active ML model is selected to generate the annotation. This proceeds in a continuous manner, as demonstrated in Figure \ref{fig:parallel_workflow}.

\begin{figure}[ht]
\centering
\includegraphics[width=0.7\linewidth]{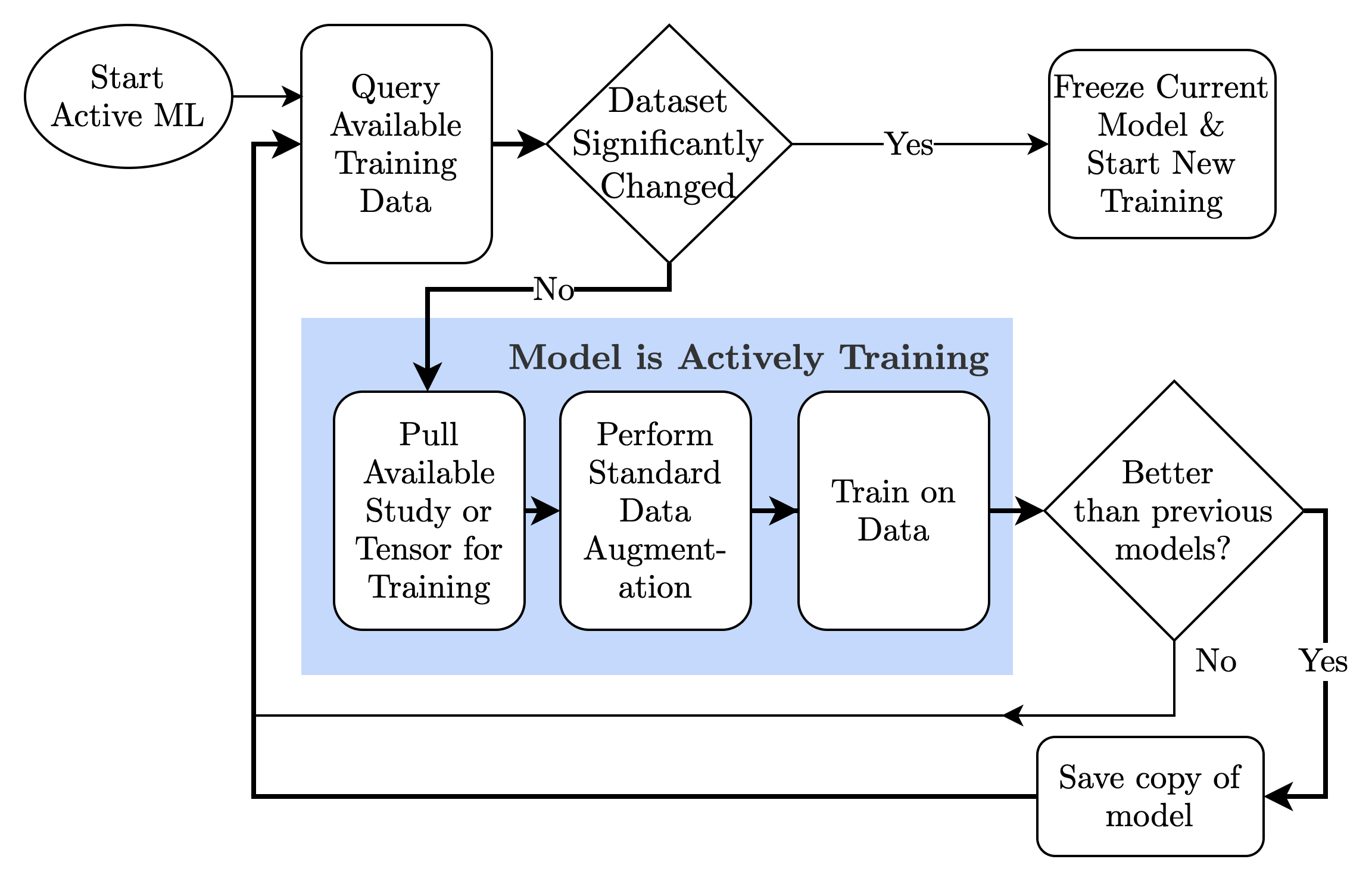}
\caption{Model Training Schematic for Continuous Active Machine Learning Strategy}
\label{fig:activeml}
\end{figure}

The system automatically pauses training as needed, or takes more dramatic action to resolve issues that can arise if the initially annotated dataset is not representative of the subsequent larger dataset, or in the case of 'catastrophic forgetting,' in which a model loses memory of previously learned information from being overwhelmed with new cases. In the former case, if there is statistically significant variation in the distribution of annotation labels, the system freezes the current model and re-initiates model training. This approach to continuous model training is shown in Figure \ref{fig:activeml}. If the model shows consistent flaws in its proposed annotations, radiologists and data scientists may evaluate and react in real-time and choose augmentations to the cohort to address these issues as they appear.

\subsubsection*{Ecosystem Integration}
As it is essential to integrate natively with pre-existing PACS infrastructure and storage, the platform includes bi-directional translation for DICOM in and DICOM out, even though the platform internally uses a structure optimized for machine interpretation. Likewise, the data and annotations must be accessible for use with common or custom development tools. The platform has an API which allows authenticated access to and integration with datasets and other components within the system. This has the added benefit of allowing third-party integration into the software platform.

Multiple instances of the platform (presumably at different institutions) can perform coordinated federating learning tasks with each other. When two or more instances share a federated learning project, model weight updates are shared between the systems so that all parties have a live copy of the model in-progress without sharing data itself. Anonymized images can optionally be shared between institutions for analysis if proper authorization is granted within the system.

\subsection{Stress-Testing the Server Instance}

Stress testing was performed by simulating concurrent users that logged in, selected an active project, and then began simulated work. Simulated users opened a new randomly selected study and series at random intervals between 5 to 10 minutes, and during each interval performed and uploaded two voxel mask annotations. These simulated annotations had the same matrix rank as the source series and contained random values. On the server, CPU, memory, and swap usage were measured. On the client systems, the latency of each request (the time until the client received a response from the server) was measured. These tests were performed for 1, 10, 20, 50, 100, 500, 1000, 5000, and 10000 simulated users.

Due to the substantial quantity of data being moved, the operations that the system handles are largely network-bound tasks. To help quantify the user experience, the time it takes to receive images from S3 storage was measured. Annotators are able to interact with images once the first image of a series has streamed in; they need not wait until the last image has arrived. Thus, the average time was measured until the first image of a series arrived, and until the entire series arrived. This measurement was performed for a simulated image series containing 200 slices of 512x512 pixel resolution, stored at 16-bit integer resolution.

Lastly, the rate at which the server could perform data ingestion from DICOM images into the ML-amenable format was measured. This was done using a test dataset containing 50 3x3 mm (slice thickness x interval) chest CT series.

\section{Results}

\begin{figure}[H]
\centering
\includegraphics[width=\linewidth]{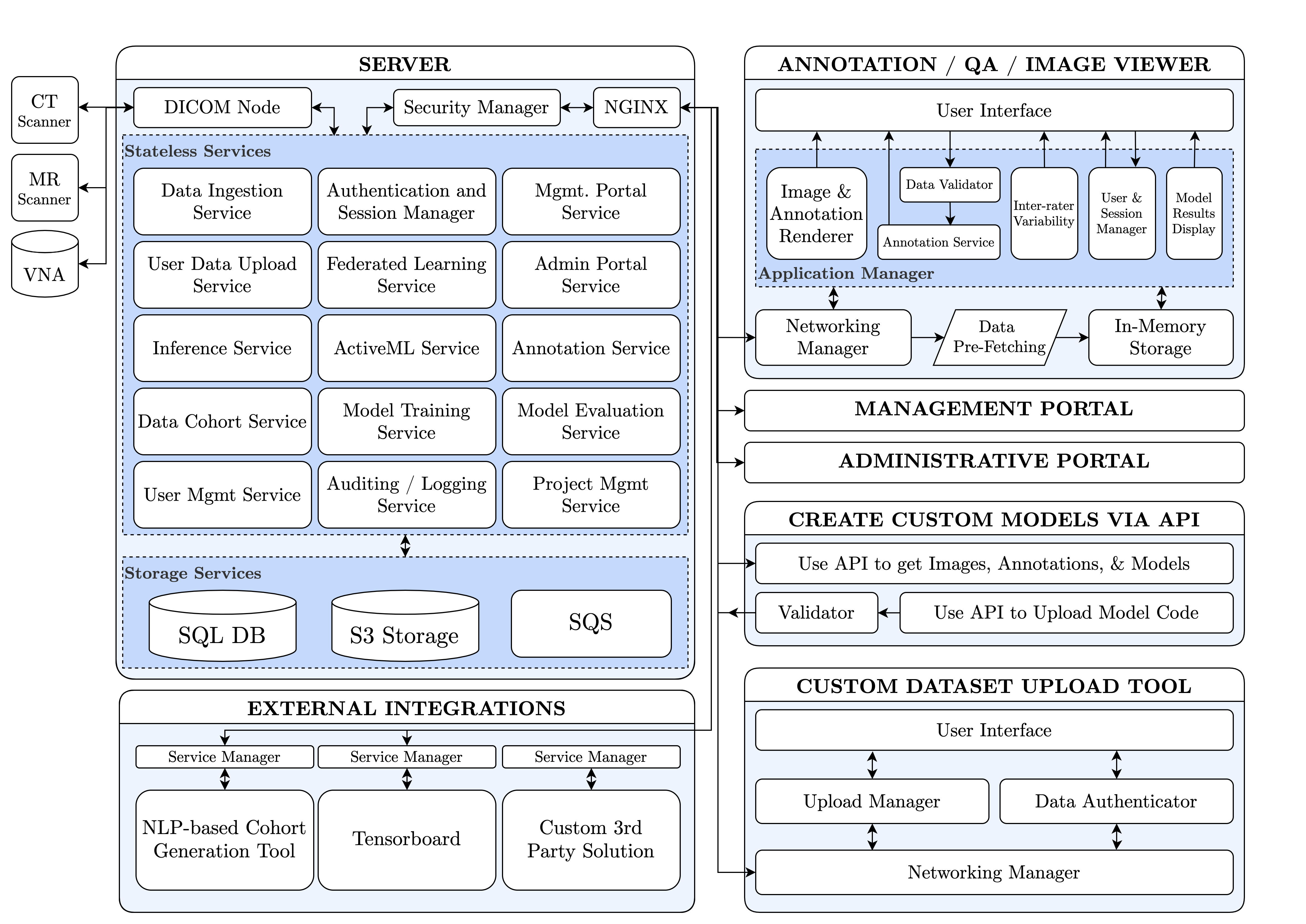}
\caption{System Architecture and Design for Tomosuite}
\label{fig:tomosuite_architecture}
\end{figure}

By strictly following the design principles detailed above, we created a machine learning platform and ecosystem which we named "Tomosuite" (paying homage to its roots in CT imaging as the Tomographic Imaging AI Suite). A high-level overview is displayed in the architecture diagram of Figure \ref{fig:tomosuite_architecture}. It includes a server module which coordinates all resources (upper left of Figure \ref{fig:tomosuite_architecture}). The annotation and QA platform, which runs on Mac, Windows, and Linux operating systems, facilitates all image viewing, annotation, and QA needs for both radiologists and data scientists (upper right of Figure \ref{fig:tomosuite_architecture}). Data, people, projects, and models are handled by the management portal, and finer control and system management is achieved through the administrative portal (middle right of Figure \ref{fig:tomosuite_architecture}). A custom-built DICOM node can query, retrieve, and pull data, and is able to push DICOM-SR objects after running model inference on an image set. An additional application enables users to upload external data and manage data cohorts under IRB protocols (bottom right of Figure \ref{fig:tomosuite_architecture}). An integrated active machine learning pipeline facilitates faster annotation and enables real-time model feedback. Tomosuite integrates with 3rd party solutions (bottom left of Figure \ref{fig:tomosuite_architecture}), like Tensorboard, for tracking model progress, and can integrate with other copies of Tomosuite at different sites to facilitate federated learning efforts. Selected components are described in greater detail below.

To step through typical usage of Tomosuite, a radiologist goes online to the management portal, registers an IRB protocol, creates a project, and adds other users to that project. This radiologist may then use the management portal to query a dataset from the hospital VNA. Another collaborator might use the data upload tool to add data that they have gathered locally on their machine. They use the annotation platform to select the data to be added to the annotation cohort and subsequently annotate that data. Throughout the project, they can track the progress of active machine learning model training on the management portal, and use it in the annotation platform when it is mature. A project PI can track the progress of project participants and generate reports through the management platform. Data scientists, working with the radiologists, can use the model development service to quickly generate models, or can use the API for external integrations with relevant components. If high level access is needed, or if something goes wrong, an authorized administrator can use the administrative portal to perform system management tasks. All the while, the server ensures that these functions are performed rapidly in a swarm-like behavior for ongoing jobs in the system.

\subsubsection*{Server Architecture}
The implementation of Tomosuite is strictly modular and containerized in Docker. Kubernetes manages these Docker instances, and the system automatically scales its resource use up or down in real-time to accommodate user and project needs, limited only by the computational resources to which it has access. Additional compute resources can be trivially added as needed, and the system will automatically expand into the new space.

The developed components and modules are shown in Figure \ref{fig:tomosuite_architecture}. The stateless services and storage services in the server module are a realization of the principles addressed in Figure \ref{fig:modular_workflow}.

\subsubsection*{Viewer and Annotation Platform}
Intuitive controls permit easy mapping of available data into research cohorts, including an interface for rapid selection of similar image types through 'smart-filtering' of standardized and previously parsed DICOM tags. With the click of a button, the backend server can generate or update a project's data cohort.

\begin{figure}[H]

\begin{subfigure}[t]{0.5\linewidth}
\includegraphics[width=\linewidth]{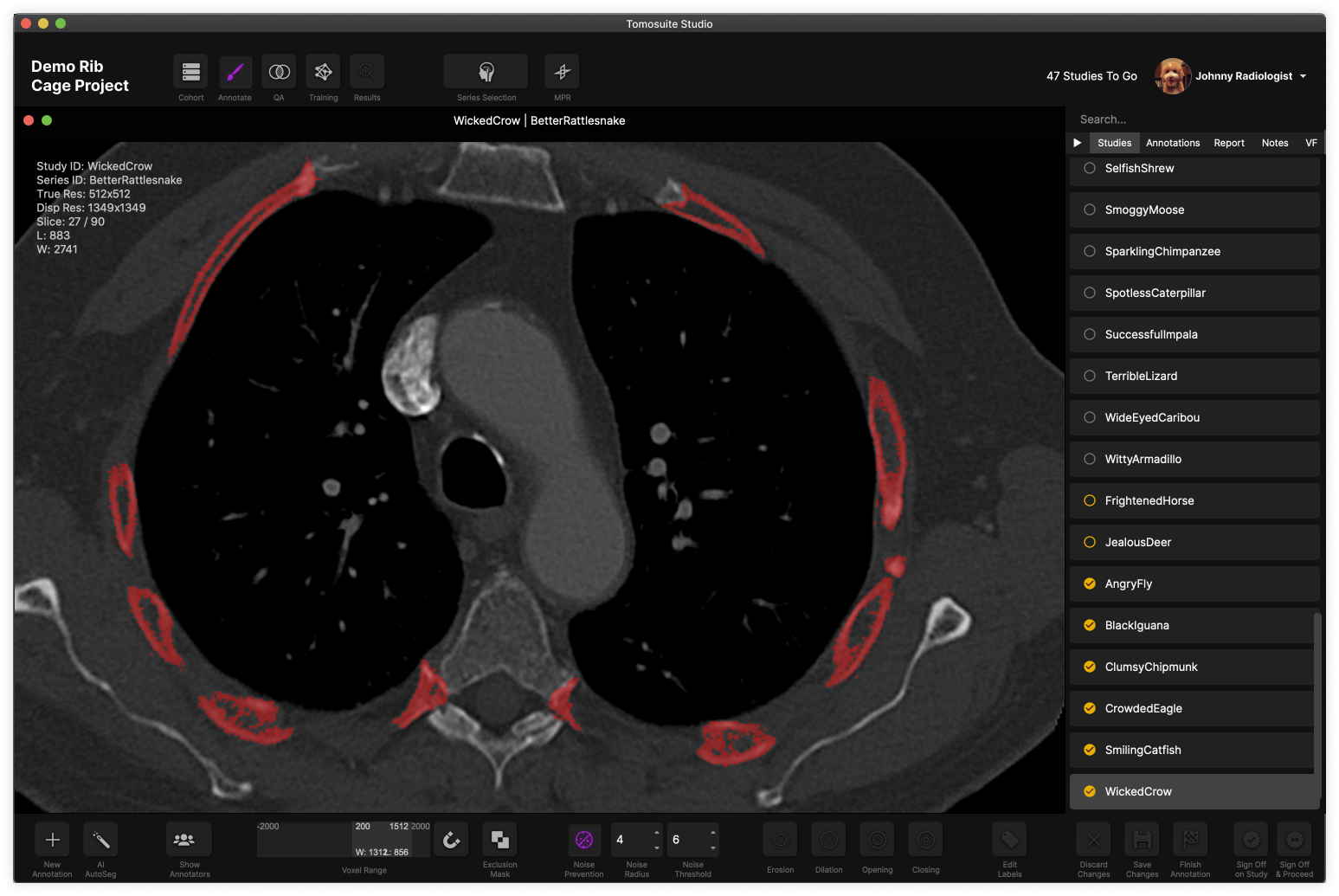} 
\caption{Annotation and Viewer Application showing annotated data generated from ML-ready data. The viewer contains controls such as windowing / leveling and other common image manipulations.}
\label{fig:Tomostudio_Viewer}
\end{subfigure}
\begin{subfigure}[t]{0.5\linewidth}
\includegraphics[width=\linewidth]{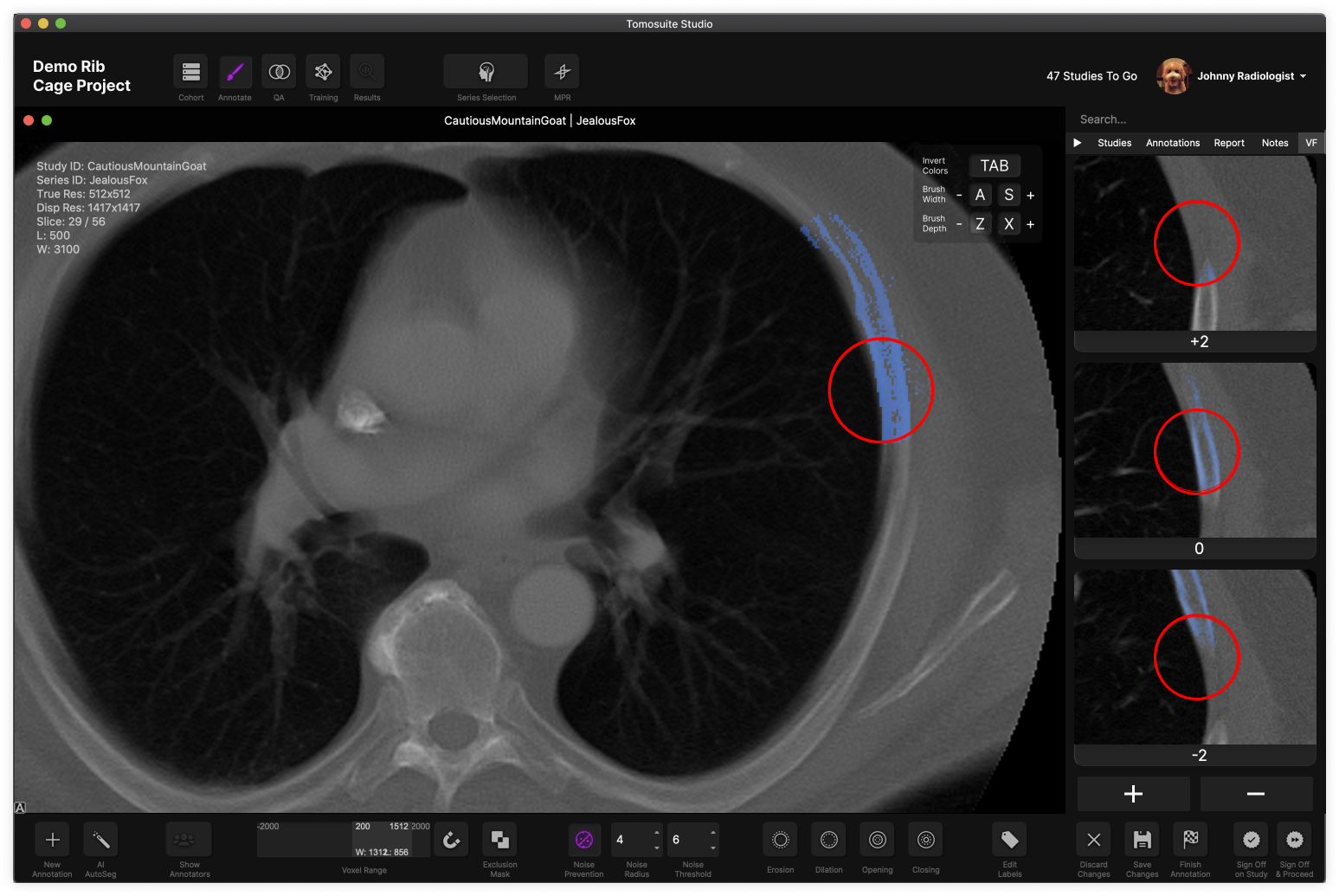}
\caption{Annotation Tool being used to place a voxel mask across multiple slices. Individual slices are arrayed to the right of the viewer.}
\label{fig:Tomostudio_VoxelMask}
\end{subfigure}
\begin{subfigure}[t]{0.5\linewidth}
\includegraphics[width=\linewidth]{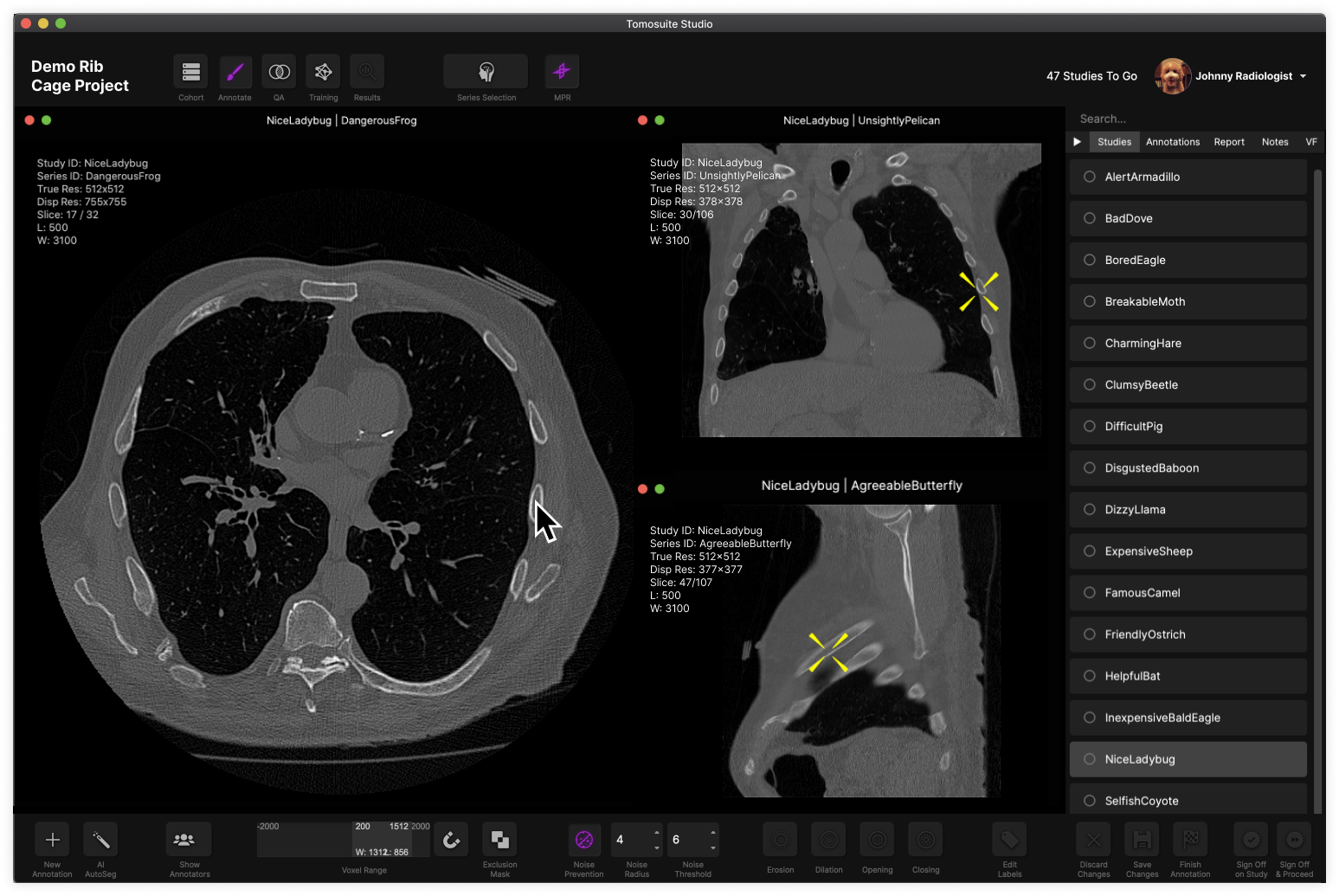}
\caption{Multiplanar Viewing. These features can also be used to translate annotations across non-standard image orientations.}
\label{fig:Tomostudio_Multiplanar}
\end{subfigure}
\begin{subfigure}[t]{0.5\linewidth}
\includegraphics[width=\linewidth]{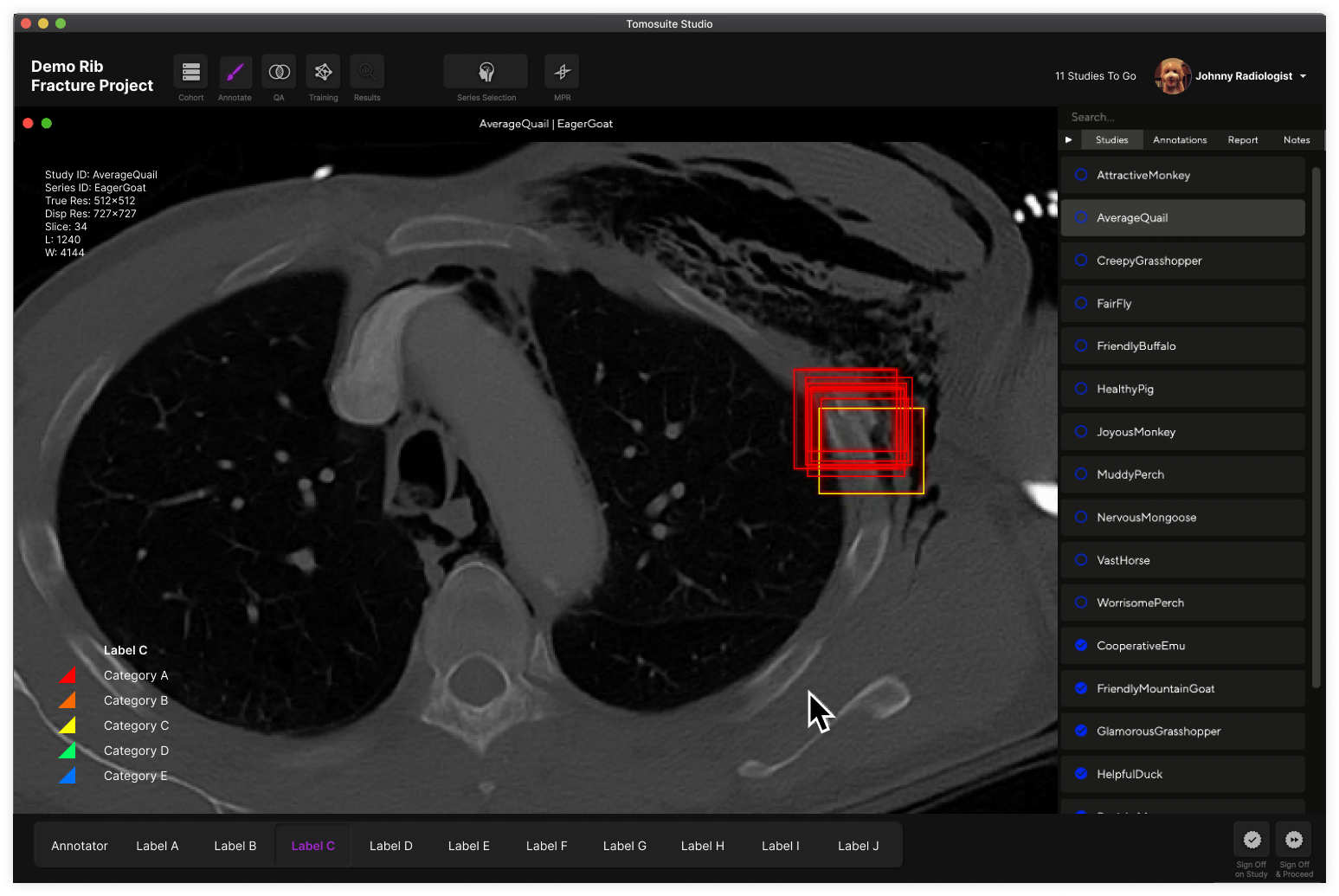}
\caption{Inter-rater variability displayed in the QA functionality in the Annotation Tool.}
\label{fig:Tomostudio_QA}
\end{subfigure}

\caption{Selected Screen-Captures from the Viewer and Annotation Software}
\label{fig:Tomostudio}
\end{figure}

The viewer displays images equivalent to a PACS, yet the source images are generated from the post-ingestion matrix objects (Figure \ref{fig:Tomostudio_Viewer}).
The viewer includes controls for the annotation tasks, such as voxel range masking (Figure \ref{fig:Tomostudio_VoxelMask}), inclusion and exclusion masks, multi-slice painting and multiplanar-enabled controls (Figure \ref{fig:Tomostudio_Multiplanar}), morphological operations, and controls to translate annotation objects between series.

Additional features to aid radiologists in their work include inline report text viewing and inter-user notes, which can only be used for communication, not model training. There are intuitive controls for multiplanar viewing, creating and managing annotations, and for viewing and interacting with the results of active machine learning models or other machine learning output as image overlays. Annotation comparison and inter-rater variability is integrated into a review panel (Figure \ref{fig:Tomostudio_QA}), enabling the platform to be used as a feedback and teaching tool for less experienced radiologists. The platform includes further optimizations such as smart data pre-fetching, which streams anticipated image stacks in the background to overcome potential challenges arising from poor network speeds or other issues that can present technical barriers for project completion.

\subsubsection*{Management Software}
\begin{figure}[H]

\begin{subfigure}[t]{0.5\linewidth}
\centering
\includegraphics[width=\linewidth]{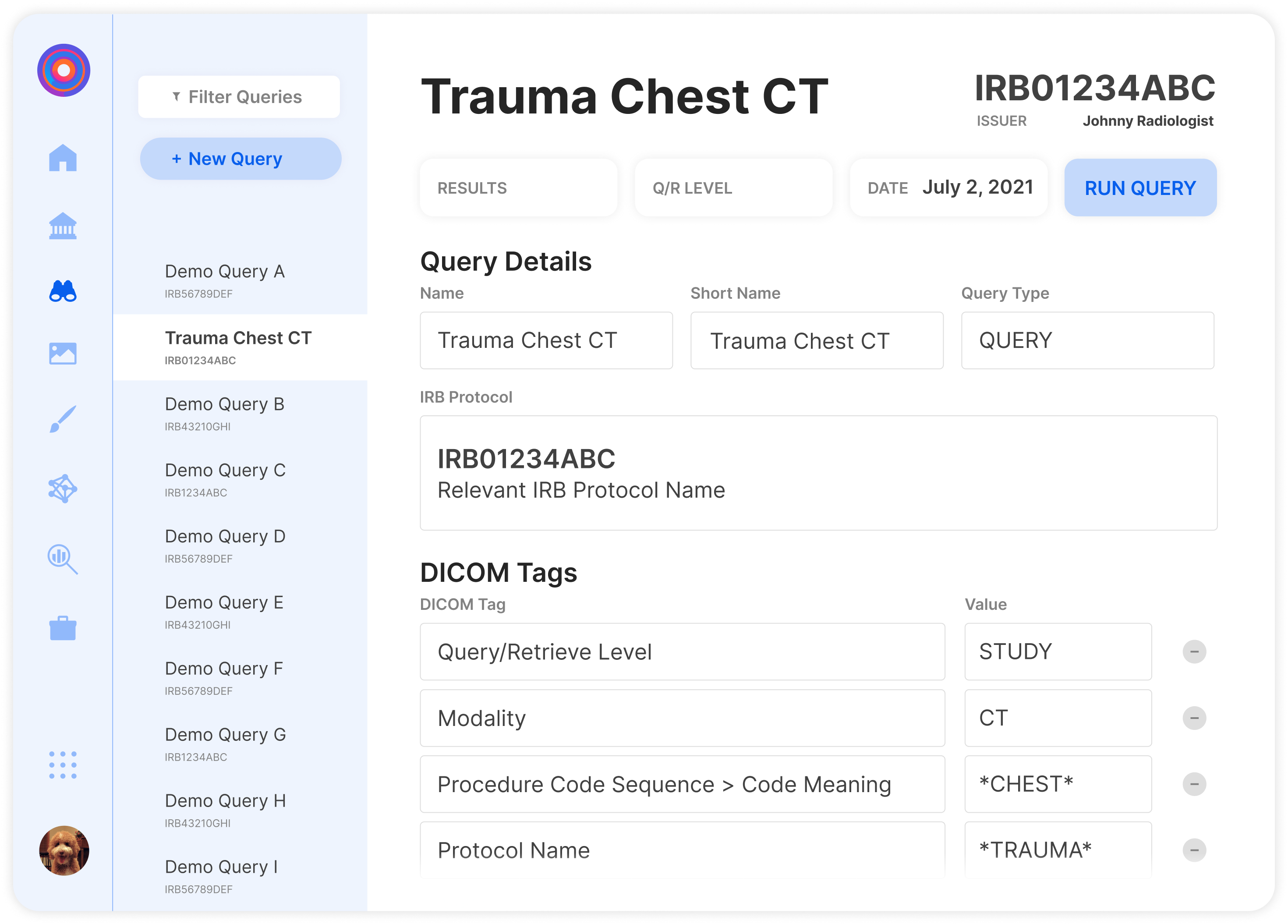} 
\caption{Management portal being used to query a VNA for available studies meeting search criteria.}
\label{fig:CentralWebsite}
\end{subfigure}
\begin{subfigure}[t]{0.5\linewidth}
\centering
\includegraphics[width=\linewidth]{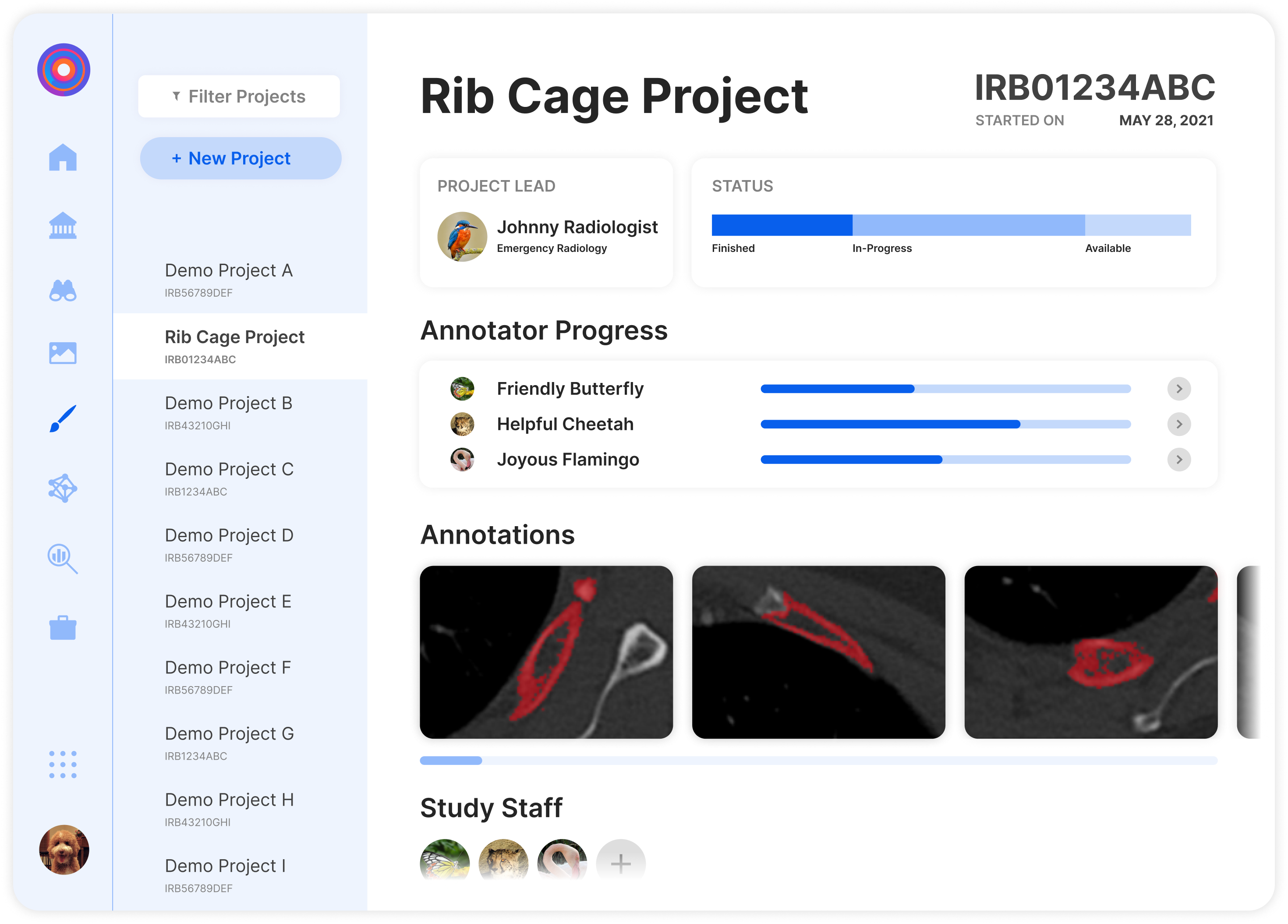}
\caption{Management portal being used to oversee an ongoing project.}
\label{fig:CentralAdmin}
\end{subfigure}

\caption{Selected Screen-Captures from the Management Portal}
\label{fig:Central}
\end{figure}

The management portal (Figure \ref{fig:Central}) facilitates all management efforts, including creating and coordinating projects, mapping to IRB protocols, overseeing users on projects, pulling data from the VNA, governing data cohorts, and tracking model progress. Once models are satisfactory, they can be deployed and monitored for inference usage. Additionally, Tomosuite includes controls to facilitate federated learning approaches.

All actions performed on both the server and within the annotation platform are logged and time-stamped. This makes governing data access, managing security, and auditing simpler to manage. Auditing data and progress reports are automatically generated, aiding in tracking time and productivity of annotators or to simplify IRB oversight.

The administrative portal gives full database and log-level control of data and records to administrators with appropriate access, and is used for any needed platform or project troubleshooting. In this way, radiologists and data scientists can manage ongoing projects without further aid or staffing.

\subsubsection*{Integrating Model Training to Run Automatically and in Parallel}
As models train, the system takes snapshots of the best versions and stores them so they can be used by the active machine learning system. This system pulls a copy of the best snapshot (according to the user-selected metric of success) of an in-process or otherwise stored model along with a copy of the requested image series, and proposes an annotation on that series. The results are stored and sent back to the requesting annotator, thereby completing the loop on the active machine learning strategy.

The parallelized machine learning implementation enables us to minimize our resource footprint. Instead of requiring expensive and high-cost computational resources, models can train at a slower pace on smaller computational resources, since models can often to train faster than radiologists can annotate. This eliminates the need for prohibitively expensive hardware resources, and instead enables users to develop preliminary models and proofs of concept on pre-existing available hardware. The system can be redirected to run on high capacity computational resources if and when they become available.

\subsubsection*{DICOM Node for Bi-Directional Communication with PACS}
The DICOM Node enables bi-directional communication with the rest of the DICOM ecosystem, pulling from and pushing to other DICOM nodes. An ingestion module converts incoming DICOM data to ML-representations, and outgoing model results back into DICOM and DICOM-SR before the DICOM node pushes them to other nodes. This approach maintains rigorous data representation and storage within the platform, while remaining harmonious with the rest of the DICOM world and leveraging DICOM for its benefits in imaging data communication.

\subsubsection*{Structured API to make Data Accessible for Advanced Analysis}
Tomosuite incorporates an API to integrate with other systems or to expand its functionality. This API enables processing and filtering of cohorts; retrieval of data and annotations; addition of models for active machine learning, model training, or deployment into the inference modules; additional federated learning tasks; and integration with other 3rd party components, such as NLP solutions or monitoring solutions like Tensorboard. Tomosuite is designed to be integrated with pre-existing infrastructure so that the system can behave both natively and with the extensibility to grow with infrastructure needs.

\subsubsection*{Evaluation}
To date and throughout its development, Tomosuite has been used for 12 projects, including slice-level qualifications, organ segmentation, and pathology detection tasks.

Before deployment to virtualized servers, Tomosuite was first deployed to common-place hardware available to our small lab during our start-up phase. The first system to run Tomosuite was an old PC with an Intel Pentium Dual E2200 @ 2x 2.2GHz CPU, 4GB DDR2 RAM, and no GPU. This system was able to support early operations during the development of the platform, even for users' computers connecting with network speeds of 18Mbps download and 8Mbps upload (well below industry standard). To run the model containers a MacBook Pro was used, which had an 8-Core i9 2.3Ghz CPU, 64GB DDR4 RAM, and Intel UHD Graphics 630 1536MB. Shortly thereafter, the entire system was hosted on this MacBook Pro computer.

Following this initial deployment, the system was migrated to an institution-provided virtual server with a 4-Core 3.00Ghz CPU, and 4GB RAM. An institution-provided S3 instance and PostgreSQL database instance were used for storage. Model training Docker containers are run as jobs on other low-cost available virtual servers.

\begin{figure}[H]

\begin{subfigure}[t]{0.33\linewidth}
\centering
\includegraphics[width=\linewidth]{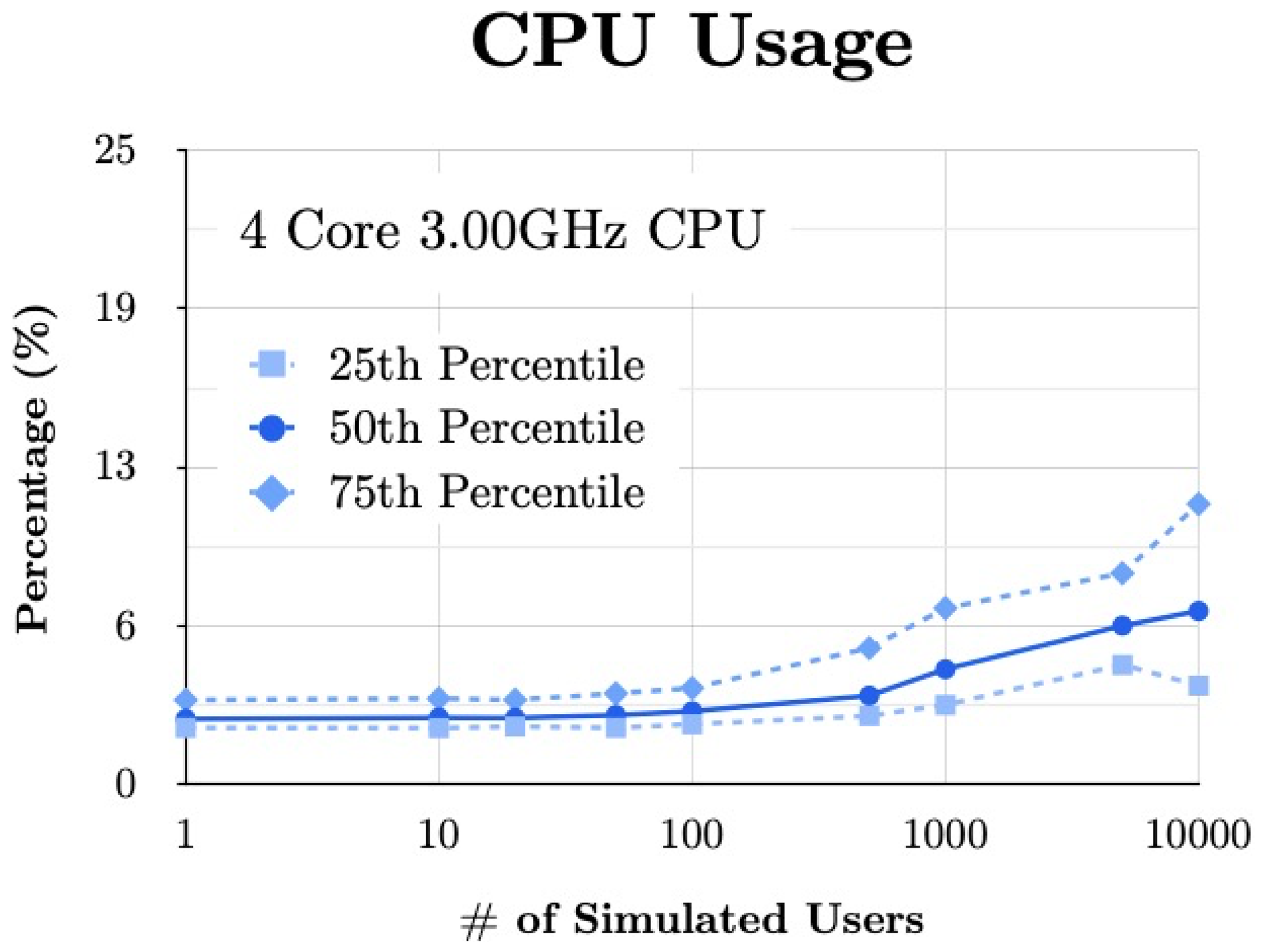}
\label{fig:TomosuiteStressTestLatency}
\end{subfigure}
\begin{subfigure}[t]{0.33\linewidth}
\centering
\includegraphics[width=\linewidth]{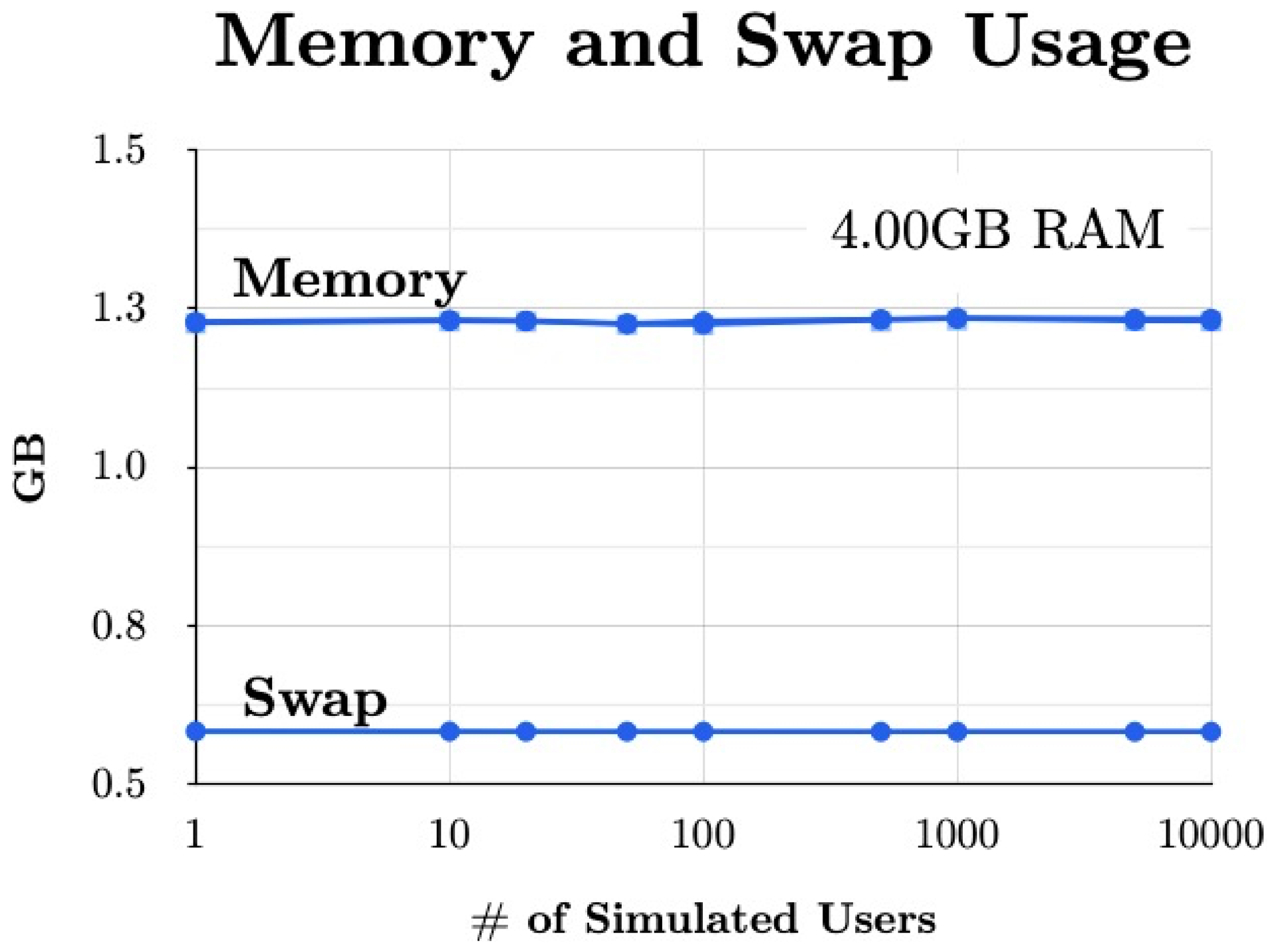}
\label{fig:TomosuiteStressTestMemory}
\end{subfigure}
\begin{subfigure}[t]{0.33\linewidth}
\centering
\includegraphics[width=\linewidth]{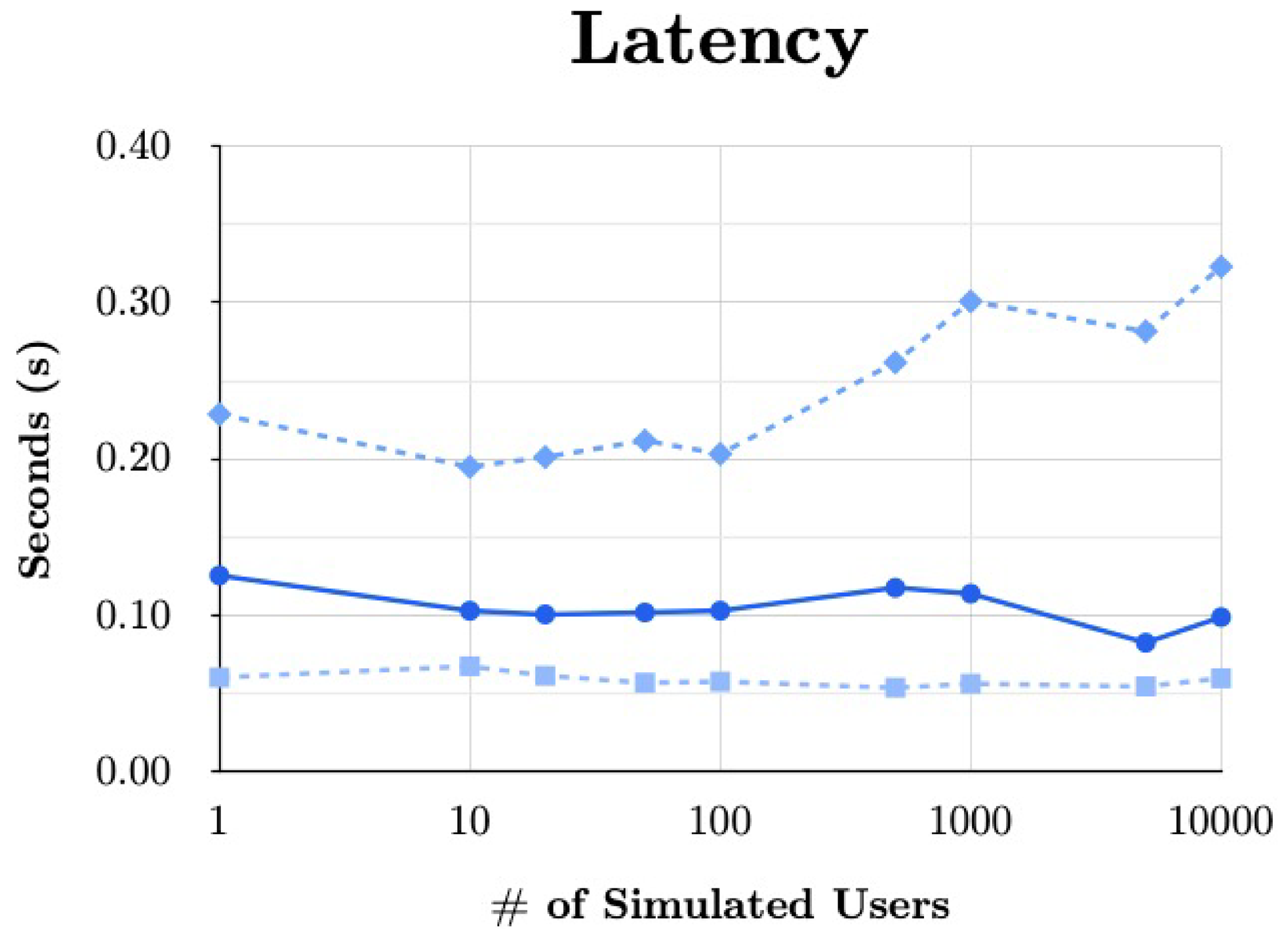}
\label{fig:TomosuiteStressTestSwap}
\end{subfigure}

\caption{Stress Test Performed on Tomosuite Server. The test was run for 1, 10, 20, 50, 100, 500, 1000, 5000, and 10000 simulated users. In memory and swap usage, the percentile lines fall closely on top of each other.}
\label{fig:TomosuiteStressTest}
\end{figure}

The simulated stress-testing of this server instance is shown in Figure \ref{fig:TomosuiteStressTest}. Simulated clients had a network speed capped at 500Mbps over VPN. Noting the logarithmic scale on the x-axis, the CPU usage appears to behave linearly, as O(n), and memory usage behaves statically, as O(1), for normal everyday usage. This is possible because the network I/O operations are offloaded to the S3 instance, and because the computational overhead is performed before data can be viewed and annotated during the ingestion step. These results demonstrate that network I/O will become a limiting factor far before CPU or memory limitations.

Measured network performance from S3 Storage, using a simulated 200-slice 512x512 pixel resolution scan, yielded 0.25 $\pm$ 0.08 seconds for the first image of a series to arrive and 9.7 $\pm$ 1.0 seconds until the entire series was available on a user's computer. This performance metric is highly subject to the network speed of users and their host institutions.

The measured data ingestion performance of the server in converting DICOM images into the ML-amenable format took 0.38 $\pm$ 0.03 seconds per image slice. In a simulated use case ingesting 50 3x3 mm chest CT series, this resulted in 28 $\pm$ 8.6 seconds per series, meaning that users or inference workflows only needed to wait approximately 30 seconds from the start of data ingestion. Processing time can be further reduced with increased compute.

\subsubsection*{Versatility for Expansion}
The system architecture has easily supported platform expansion. New features and functionality are integrated by adding individual modules into the zoo of stateless services in the server module. This has allowed us to respond to the evolving needs of radiologists and data scientists on various projects.

\section{Discussion}
Despite recent advances, many components of traditional AI workflows still require significant resources and labor, often necessitating large teams to overcome data incongruities and workflow gaps. These teams require broad interdisciplinary expertise in data science and artificial intelligence, radiology, and computer software and engineering, in order to diagnose and troubleshoot the issues that inevitably arise.

Our efforts focused on creating a system that maximized computer-assisted automation, and minimized human resources and wasted time. Importantly, radiologist users do not need advanced computer experience or significant resources to carry out AI research. Tomosuite's ultimate efficiency and high workflow automation result from meticulously addressing the myriad insidious problems that collectively have the power to halt AI research projects. The solutions to many of these problems share common themes: attention to data format, cloud-oriented algorithmic structure, workflow re-organization, interoperability, and optimizations enabled by a high degree of workflow parallelization.

Radiology systems and standards, such as PACS and DICOM, have been largely clinician oriented, designed to streamline the process by which radiologists perform clinical evaluations. While these effectively ushered in the digitization of medical imaging, they were not designed to meet challenges common to the era of artificial intelligence. Computer systems require uniform standardization, rigorous accuracy, and in the case of machine learning, large quantities of data in order to train models. Radiologists can trivially recognize axial from coronal, or a scout image from the rest of an image stack, or even overlook incorrect data in a metadata tag or fill in missing data with contextual reasoning. However, these issues are often show-stopping for computer algorithms. DICOM and DICOM-SR are communication formats that vary from vendor to vendor, without guarantees of standardization of data representation or metadata. Rather than retrofitting machine learning solutions to use these formats at every stage of the pipeline, our system formats and stores data in matrix, vector, and otherwise uniformly standardized formats. This enables a maximization of automated computer management, processing, and analysis, which saves time for both radiologists and data scientists. Indeed, the resources and staffing needed to facilitate and troubleshoot long and arduous ingestion efforts are no longer required when data and annotations are handled in this fashion.

This shift in data format has many benefits for the development of modules and algorithms within the platform. Clean, standardized, ML-ready data is used for all internal modules in the system, reducing the overhead needed to develop additional workflow components. Continued development on workflow modules is concerned only with the objectives of their respective algorithms, and is not hindered by challenges that can arise elsewhere in the pipeline, such as the ingestion process. Furthermore, by building modules without cross-dependencies or even the need to communicate directly with other processes, the system is able to perform in cloud-computing environments and rapidly grow in scope and scale. For data scientists and developers, the system’s modular design has the advantage that components can be used as they become available, without concern about creating cross-dependencies and proverbial "spaghetti code" that can plague project developments and necessitate long, manual interventions. As the platform expands, older modules need not be revisited as their function and implementations are self-contained. Completed modules collectively form a library of components available to all users.

Our workflow re-ordering minimizes bottlenecks, risks of data loss, and subsequent needs for recovery. The most expensive part of our pipeline (annotation) occurs after the most error-prone component (ingestion). Done this way, annotators are easily able to spot any ingestion errors that may have slipped through the cracks, providing a natural data quality assurance step that otherwise would not exist. This further ensures confidence in the dataset when model training begins. Aside from ongoing platform development, this workflow reorganization eliminates the need for data engineers to convert unstructured or ill-maintained data into regularized and usable troves.

Another focus was on building a platform that was harmonious with the broader DICOM-driven environment while responding to the needs of AI model development. The system can be integrated with other systems and solutions within the larger radiology ecosystem. The platform leverages DICOM for its external communication,  thereby enabling greater interoperability and lowering translational challenges.

The parallelization enabled by all of these strategies makes Tomosuite well-suited for the benefits of active machine learning, which has the potential to substantially reduce needed annotation effort. This approach empowers interactivity between the radiologist and model, much as an attending radiologist might interact with a resident. This strategy even has the potential to lower the data requirement for model training as each annotation individually contributes necessary medical concepts to the training cohort, including those selected to mitigate shortcomings found in earlier model iterations. It is worth noting the importance of contextualizing active machine learning functionality as an annotator-assistance tool and not an annotator-replacement tool. Careful consideration of the use of active machine learning is warranted, most especially on projects with rare or challenging findings. If a less experienced annotator trusts active machine learning suggestions against their better judgement, they can be biased into making errors they otherwise would not. With this in mind, annotators must be informed that they are ultimately responsible for the annotations they sign off on in the dataset, and to use the functionality only as an assistance tool. Used responsibly, this functionality can provide dramatic reductions to annotation time.

We have been able to provide a clean and simple user experience while ensuring high rigor in the collection and maintenance of data. Our system is intended to be used comfortably and efficiently by experienced radiologists, while handling data in an ML-optimized and cloud-computing oriented manner. Ultimately, all of these many areas of focus come together to form a platform that can facilitate imaging AI research in a highly efficient, and resource-minimizing fashion.

\section{Conclusion}

The described methodology and platform originated from the desire to perform meaningful AI research without the benefit of a well-populated AI lab or a huge up-front investment in technical resources. The up-front conversion to an ML-ready data format, and the segregation of ecosystem functionality to modular components has enabled a platform that is not only rich in functionality, but can also run on low-cost compute and meet the needs of groups like ours that wish to be productive and nimble despite limited initial resources. We built a research pipeline and infrastructure solution to meet our needs, but believe that these design principles and approaches may similarly benefit other resource-limited labs, or alternatively, may greatly enhance the efficiency of larger groups and resource-rich institutions.

To accomplish these objectives, it was necessary to return to the fundamentals and take a hard look at the source and quality of our data. The question is not only 'How can we build models to fit the data?' but also 'How can we improve the data to create better models?' Our approach focused on designing an AI development pipeline that meets radiologist's ease of use and efficiency needs, while adhering to rigorous data science requirements. This infrastructure design has the potential to greatly accelerate AI research by sidestepping many of the time consuming tasks that typically consume the attention of AI teams. In doing so, AI can become another tool that radiologists can use to drive innovation that advances patient care. 
\clearpage

\section*{Compliance with Ethical Standards}

\subsection*{Ethics approval}
IRB approval was obtained with waiver of informed consent for retrospective medical record review to obtain the imaging datasets used in the development and evaluation of the platform.

\section*{Acknowledgements}
The authors are grateful to Samantha Bloom for her work in developing and elevating the graphical design of the Tomosuite platform and to Sam Stern for many thoughtful conversations about server design and infrastructure.

\bibliographystyle{unsrt}
\bibliography{bibliography}

\end{document}